\documentclass[12pt]{elsart}

\usepackage{t1enc}
\usepackage[latin1]{inputenc}
\usepackage[english]{babel}

\usepackage{graphicx}

\journal{Nucl. Instr. and Meth. A}

\def\baselinestretch{1.4}

\begin{document}

\begin{frontmatter}

\title{Space-frequency model of amplified spontaneous emission and super-radiance
in free electron laser operating in the linear and non-linear
regimes}

\author{Yosef Pinhasi\thanksref{email}},
\thanks[email]{\em E-mail adress: yosip@eng.tau.ac.il}
\author{Yuri Lurie},
\author{Asher Yahalom},
and
\author{Amir Abramovich}
\address{
The College of Judea and Samaria,
Dept. of Electrical and Electronic Engineering --- Faculty of Engineering,
P.O. Box 3, Ariel 44837, Israel }

\maketitle

\begin{abstract}

A three-dimensional, space-frequency model for the excitation of
electromagnetic radiation in a free-electron laser is presented.
The approach is applied in a numerical particle code WB3D,
simulating the interaction of a free-electron laser operating in
the linear and non-linear regimes. Solution of the electromagnetic
excitation equations in the frequency domain inherently takes into
account dispersive effects arising from the cavity and the gain
medium. Moreover, it facilitates the consideration of statistical
features of the electron beam and the excited radiation, necessary
for the study of spontaneous emission, synchrotron amplified
spontaneous emission (SASE), super-radiance and noise.

We employ the code to study the statistical and spectral characteristics of the
radiation generated in a pulsed beam free-electron laser operating in the
millimeter wavelengths. The evolution of radiation spectrum, excited when a
Gaussian shaped bunch with a random distribution of electrons is passing
through the wiggler, was investigated. Numerical results of spontaneous
emission power along the wiggler are compared to analytical predictions in the
linear regime. In the first few periods, the power is excited from shot-noise
in the low-gain regime. An exponential growth of SASE in the high-gain regime
is inspected after passing a sufficient number of periods, until saturation
occurs when arriving to the non-linear regime of the FEL operation.

\end{abstract}

\end{frontmatter}


\section{Introduction}

Electron devices such as microwave tubes and free-electron lasers
(FELs) utilize distributed interaction between an electron beam
and electromagnetic radiation. Random electron distribution in the
$e$-beam causes fluctuations in current density, identified as
{\em shot noise} in the beam current
\cite{Schottky}-\cite{Smulin59}.
Electrons passing through
a magnetic undulator emit a partially coherent radiation which is
called {\em undulator synchrotron radiation} \cite{3.4}. The
electromagnetic fields excited by each electron add incoherently,
resulting in {\em spontaneous emission}
\cite{3.6}-\cite{3.11}.
When the electron beam is modulated or pre-bunched, the fields excited by
electrons become correlated, and coherent summation of radiation fields from
individual particles occurs. If all electrons radiate in phase with each other,
the generated radiation becomes coherent ({\em super-radiant emission}).

In high-gain FELs, utilizing sufficiently long undulators, the
spontaneous emission radiation excited in the first part of the
undulator is amplified along the reminder of the interaction
region resulting in self-amplified spontaneous emission (SASE)
\cite{new15}-\cite{new19}.
Super-radiant emission emerges if the electrons are injected
into the undulator in a single short bunch
(shorter than the oscillation period of the emitted radiation)
\cite{3.12}-\cite{3.21b}
or enter as a periodic train of bunches at the frequency of the emitted radiation
\cite{3.18}-\cite{SR_2000}.

Investigation of spontaneous and super-radiant emissions, as well
as SASE, call for analytical and numerical models, that can
describe non-stationary stochastic processes involved in the
generation of incoherent or partially coherent radiation. In
addition to the statistical features of the electron beam and
radiation, the model should take into account dispersive effects
evolving in the medium, in which the radiation is excited. The
model presented in this paper, utilizes an expansion of the total
electromagnetic field (radiation and space-charge waves) in terms
of transverse eigenmodes of the medium, in which the field is
excited and propagates \cite{FEL3D_description,WBI}. The
interaction between the electron beam and the electromagnetic
field is described by a set of coupled-mode excitation equations,
that expresses the evolution of mode amplitudes in the frequency
domain. The excitation equations are solved self-consistently with
the equation of particles motion, which describes the electron
beam dynamics.

\section{Presentation of the electromagnetic field in the frequency domain}

The electromagnetic field in the time domain is described by the
space-time electric ${\bf E}({\bf r},t)$ and magnetic ${\bf
H}({\bf r},t)$ signal vectors. ${\bf r}$ stands for the $(x,y,z)$
coordinates, where $(x,y)$ are the transverse coordinates and $z$
is the axis of propagation. The Fourier transform of the electric
field is defined by:

\begin{equation}
{\bf E}({\bf r},f) = \int_{-\infty}^{+\infty} {\bf E}({\bf r},t) \ e^{+j 2 \pi f t} \ d f
\end{equation}
where $f$ denotes the frequency. Similar expression is defined for
the Fourier transform ${\bf H}({\bf r},f)$ of the magnetic field.
Since the electromagnetic signal is real (i.e. ${\bf
E}^{\ast}({\bf r},t)={\bf E}({\bf r},t)$), its Fourier transform
satisfies ${\bf E}^{\ast}({\bf r},f)={\bf E}({\bf r},-f)$.

Analytic representation of the signal is given by the complex
expression:

\begin{equation}
\widetilde{\bf E}({\bf r},t) \equiv {\bf E}({\bf r},t) - j \, \widehat{{\bf E}({\bf r},t)}
\label{analytic}
\end{equation}
where

\begin{equation}
\widehat{{\bf E}({\bf r},t)} = \int_{-\infty}^{+\infty} \frac{{\bf E}({\bf r},t') } {t-t'} \ d t'
\end{equation}
is the Hilbert transform of ${\bf E}({\bf r},t)$. Fourier
transformation of the analytic representaion (\ref{analytic})
results in a 'phasor-like' function $\widetilde{\bf E}({\bf r},f)$
defined in the positive frequency domain and related to the
Fourier transform by:

\begin{equation}
\widetilde{\bf E}({\bf r},f) = 2 \, {\bf E}({\bf r},f) \cdot u(f)
= \left\{
\begin{array}{cl}
2 \, {\bf E}({\bf r},f) \ \ \  & f > 0 \\
0  & f < 0
\end{array}
\right.
\label{phasorlike}
\end{equation}
The Fourier transform can be decomposed in terms of the 'phasor
like' functions according to:

\begin{equation}
{\bf E}({\bf r},f) = \frac{1}{2} \, \widetilde{\bf E}({\bf r},f) +
\frac{1}{2} \, \widetilde{\bf E}^\ast({\bf r},-f)
 \label{composite}
\end{equation}
and the inverse Fourier transform is then:

\begin{equation}
{\bf E}({\bf r},t) =
\int_{-\infty}^{+\infty} {\bf E}({\bf r},f) \ e^{-j 2 \pi f t} \ d f
= \Re \left\{
\int_{0}^{\infty} \widetilde{\bf E}({\bf r},f) \ e^{-j 2 \pi f t} \ d f \right\}
\label{Eq(r,t)1}
\end{equation}

\section{The Wiener-Khinchine and Parseval theorems for electromagnetic fields}

The cross-correlation function of the time dependent electric
${\bf E}({\bf r},t)$ and magnetic ${\bf H}({\bf r},t)$ fields is
given by:

\begin{equation}
R_{EH}(z,\tau)=
\int_{-\infty}^{+\infty}
\left\{ \int\int
\left[ {\bf E}({\bf r},t+\tau) \times {\bf H}({\bf r},t) \right]
\cdot \hat{\bf z} \ d x \, d y \right\} \ d t
\label{correlation}
\end{equation}
Note that for finite energy signals, the total energy carried by
the electromagnetic field is given by ${\mathcal W}(z)= R_{EH}(z,0)$

According to the Wiener-Khinchine theorem, the spectral density
function of the electromagnetic signal energy $S_{EH}(z,f)$ is
related to the Fourier transform of the cross-correlation function
$R_{EH}(z,\tau)$ through the Fourier transformation:

\begin{eqnarray}
\mbox{} \hspace{-5mm}
S_{EH}(z,f) &=& \int_{-\infty}^{+\infty} R_{EH}(z,\tau) \ e^{+j 2\pi f \tau} \  d \tau
= \int\int \left[ {\bf E}({\bf r},f) \times {\bf H}^\ast({\bf r},f) \right] \cdot \hat{\bf z}
\ d x \, d y
=  \nonumber \\
&=& \left\{
   \begin{array}{ll}
       \int \int \frac{1}{4} \left[
          \widetilde{\bf E}({\bf r},f) \times \widetilde{\bf H}^\ast({\bf r},f) \right]
             \cdot \hat{\bf z} \ d x \, d y & f>0 \\
       \int \int \frac{1}{4} \left[
          \widetilde{\bf E}({\bf r},-f) \times \widetilde{\bf H}^\ast({\bf r},-f)\right]^\ast
             \cdot \hat{\bf z} \ d x \, d y \ \ \  & f<0
 \end{array}  \right.
\end{eqnarray}

Following Parseval theorem, the total energy carried by the
electromagnetic field can also be calculated by integrating the
spectral density $S_{EH}(z,f)$ over the entire frequency domain:

\begin{eqnarray}
 {\mathcal W}(z) &=& \int_{-\infty}^{+\infty} S_{EH}(z,f) \ d f =  \nonumber \\
 &=& 
\int_0^\infty \left[
\int\int \frac{1}{2} \ \Re \left\{
 \widetilde{\bf E}({\bf r},f) \times \widetilde{\bf H}^\ast({\bf r},f) \right\}
           \cdot \hat{\bf z} \ d x \, d y \right] \ d f
\end{eqnarray}
We identify:

\begin{equation}
\frac{d {\mathcal W}(z)}{d f} = \frac{1}{2} \  \Re \left\{
\int\int \left[
\widetilde{\bf E}({\bf r},f) \times \widetilde{\bf H}^\ast({\bf r},f) \right]
      \cdot \hat{\bf z} \ d x \, d y \right\}
\label{esd}
\end{equation}
as the spectral energy distribution of the electromagnetic field (over positive frequencies).

\section{Modal presentation of electromagnetic field in the
frequency domain}

The 'phasor like' quantities defined in (\ref{phasorlike}) can be
expanded in terms of transverse eigenmodes of the medium in which
the field is excited and propagates \cite{WBI}. The perpendicular
component of the electric and magnetic fields are given in any
cross-section as a linear superposition of a complete set of
transverse eigenmodes:

\begin{eqnarray}
\widetilde{\bf E}_\perp({\bf r,f}) &=&
\sum_q  \left[ C_{+q}(z,f) \, e^{+ j k_{zq} z} \ +\   C_{-q}(z,f) \, e^{- j k_{zq} z} \right]
   \, \widetilde{\bf{\mathcal E}}_{q \perp}(x,y)      \nonumber  \\
\widetilde{\bf H}_\perp({\bf r,f}) &=&
\sum_q  \left[ C_{+q}(z,f) \, e^{+ j k_{zq} z} \ -\   C_{-q}(z,f) \, e^{- j k_{zq} z} \right]
   \, \widetilde{\bf{\mathcal H}}_{q \perp}(x,y)
\label{transverse1}
\end{eqnarray}
$C_{+q} (z,f)$ and $C_{-q} (z,f)$ are scalar amplitudes of the
$q$th forward and backward modes respectively with electric field
$\widetilde{\bf{\mathcal E}}_{q \perp}(x,y)$ and magnetic field
$\widetilde{\bf{\mathcal H}}_{q \perp}(x,y)$ profiles and axial
wavenumber:

\begin{equation}
k_{zq} = \left\{
\begin{array}{lll}
j \sqrt{k^2_{\perp q} - k^2} \ \ \   & k < k_{\perp q} & \mbox{(cut-off modes)} \\
\sqrt{k^2 - k^2_{\perp q}}           & k > k_{\perp q} & \mbox{(propagating modes)}
\end{array}
\right.
\label{kzq}
\end{equation}

Expressions for the longitudinal component of the electric and
magnetic fields are obtained after substituting the modal
representation (\ref{transverse1}) of the fields into Maxwell's
equations, where source of electric current density
$\widetilde{\bf J}({\bf r},f)$ is introduced:

\begin{eqnarray}
&&\mbox{} \hspace{-10mm}
\widetilde{E}_z({\bf r},f) =
\sum_q  \left[ C_{+q}(z,f)\, e^{+ j k_{zq} z} -  C_{-q}(z,f) \, e^{- j k_{zq} z} \right]
\, \widetilde{\mathcal E}_{q z}(x,y)
 +  \frac{1}{j 2 \pi f \epsilon_0} \widetilde{J}_z({\bf r},f)     \nonumber  \\
&&\mbox{} \hspace{-10mm}
\widetilde{H}_z({\bf r},f) =
\sum_q  \left[ C_{+q}(z,f) \, e^{+ j k_{zq} z} +  C_{-q}(z,f) \, e^{- j k_{zq} z} \right]
\, \widetilde{\mathcal H}_{q z}(x,y)
    \label{longitudinal1}
\end{eqnarray}

The evolution of the amplitudes of the excited modes is described
by a set of coupled differential equations:

\begin{eqnarray}
\lefteqn{\frac{d}{d z} C_{\pm q}(z,f) =} &&   \nonumber  \\
&=&  \mp \frac{1}{2 {\mathcal N}_q} \ e^{\mp j k_{z q} z}
\int \int
 \left[ \left( \frac{Z_q}{Z_q^\ast} \right) {\widetilde{\bf J}}_\perp({\bf r},f) \pm
 \hat{\bf z} \widetilde{J}_z({\bf r},f) \right]
 \cdot {\widetilde{\mathcal E}_q}^\ast (x,y) \  d x \, d y
                              \label{excitation_Coefficients}
\end{eqnarray}
The normalization of the field amplitudes of each mode is made via
each mode's complex Poynting vector power:

\begin{equation}
{\mathcal N}_q = \int \int \left[ \widetilde{\bf{\mathcal E}}_{q \perp}(x,y)
     \times \widetilde{\bf{\mathcal H}}_{q \perp}^\ast (x,y) \right]
                         \cdot \hat{\bf z} \ d x \, d y
\label{Nq}
\end{equation}
and the mode impedance is given by:

\begin{equation}
Z_q = \left\{
\begin{array}{ll}
\sqrt{ \frac{\mu_0}{\epsilon_0} } \frac{k}{k_{zq}} = \frac{2 \pi f \mu_0}{k_{zq}} \ \ \
             & \mbox{for TE modes} \\
\sqrt{ \frac{\mu_0}{\epsilon_0} } \frac{k_{zq}}{k} = \frac{k_{zq}}{2 \pi f \epsilon_0}
             & \mbox{for TM modes}
\end{array}\right.
\end{equation}

Substituting the expansion (\ref{transverse1}) in (\ref{esd})
results in an expression for the spectral energy distribution of
the electromagnetic field (over positive frequencies) as a sum of
energy spectrum of the excited modes:

\begin{eqnarray}
\frac{d {\mathcal W}(z)}{d f} &=&
\sum_{\stackrel{q}{Propagating}}
 \frac{1}{2} \left[ |C_{+q}(z,f)|^2 - |C_{-q}(z,f)|^2 \right]
 \Re \left\{ {\mathcal N}_q \right\} +
             \nonumber \\
&& \hspace{3cm}
 + \sum_{\stackrel{q}{Cut-off}} \Im \left\{ C_{+q}(z,f) C_{-q}^\ast(z,f)
\right\} \Im \left\{{\mathcal  N}_q \right\}
\label{power2}
\end{eqnarray}
The power spectral density carried by the propagating mode $q$
during a temporal period $T$ is given by the ensemble average:

\begin{equation}
\frac{d P_q(z)}{d f} = \frac{1}{T} \ \overline{ \left| C_q(z,f) \right|^2 }
                   \ {\mathcal P}_q
\label{Wq}
\end{equation}
where ${\mathcal P}_q = \frac{1}{2} \Re \left\{ {\mathcal N}_q \right\}$.

\section{The electron beam dynamics}

The state of the particle $i$ is described by a six-components
vector, which consists of the particle's position coordinates
${\bf r}_i = (x_i,y_i,z_i)$ and velocity vector ${\bf v}_i$. The
velocity of each particle, in the presence of electric ${\bf
E}({\bf r},t)$ and magnetic ${\bf B}({\bf r},t) = \mu_0 {\bf
H}({\bf r},t)$  fields, is found from the Lorentz force equation:

\begin{equation}
\frac{d {\bf v}_i}{d z} = \frac{1}{\gamma_i} \left\{- \,
\frac{e}{m} \, \frac{1}{v_{z_i}} \left[ {\bf E}({\bf r}_i,t) +
{\bf v}_i \times {\bf B}({\bf r}_i,t) \right]
  -  {\bf v}_i \, \frac{d \gamma_i}{d z} \right\}
                       \label{force_velocity}
\end{equation}
where $e$ and $m$ are the electron charge and mass respectively.
The fields represent the total (DC and AC) forces operating on the
particle, and include also the self-field due to space-charge. The
Lorentz relativistic factor $\gamma_i$ of each particle is found
from the equation for kinetic energy:

\begin{equation}
\frac{d\gamma_i}{d z} =
- \, \frac{e}{m c^2} \, \frac{1}{v_{z_i}} \ {\bf v}_i \cdot {\bf E}({\bf r}_i,t)
\label{force_gamma}
\end{equation}
where $c$ is the velocity of light.

The time it takes a particle to arrive at a position $z$, is a
function of the time $t_{0i}$ when the particle entered at $z=0$,
and its instantaneous longitudinal velocity $v_{zi}(z)$ along the
path of motion:
\begin{equation}
t_i(z)=t_{0_i} + \int_0^z \frac{1}{v_{z_i}(z')} \ d z' \label{time}
\end{equation}

The current distribution is determined by the position and the
velocity of the particles in the beam:

\begin{equation}
{\bf J} ({\bf r},t) = - q \sum\limits_{i=1}^{N}
  \left(\frac{{\bf v}_i}{v_{z_i}}\right) \, \delta (x-x_i) \ \delta (y-y_i) \,
  \delta \left[t - t_i(z)\right]
\end{equation}
here $q=\frac{I_0 T}{N}$ is the charge of each of the $N$ macro
particles in the simulation ($I_0$ is the DC current of the e-beam
pulse of temporal duration $T$). The 'phasor like' current density
is given by:

\begin{eqnarray}
{\widetilde{\bf J}}({\bf r},f) &=&
2 \, u(f) \int\limits_{-\infty}\limits^{+\infty}
       {\bf J}({\bf r},t) \  e^{-j 2 \pi f t} \ d t  =   \nonumber \\
&=&
- 2  q \sum\limits_{i=1}^{N} \left( \frac{{\bf v}_i}{v_{z_i}} \right)
     \, \delta (x-x_i) \, \delta (y-y_i) \ e^{-j 2\pi  f t_i(z)} \ u(f)
\label{current_Fourier}
\end{eqnarray}

\section{Numerical results}

The WB3D code was used to investigate the excitation of
spontaneous emission in a millimeter wave free-electron maser
(FEM), with operational parameters given in Table 1. The
corresponding dispersion curves of the FEM are shown in Fig. 1.
When the beam energy is set to 1.375~MeV, there are two separated
intersection points between the beam and waveguide dispersion
curves, corresponding to the ``slow'' ($v_{g_1} < v_{z_0}$) and
``fast'' ($v_{g_2} > v_{z_0}$) synchronism frequencies 29~GHz and
100~GHz, respectively. Lowering the beam energy to 1.066~MeV,
results in a single intersection at 44~GHz ("grazing limit"),
where the beam dispersion line is tangential to the waveguide
dispersion curve ($v_{g} = v_{z_0}$).

The evolution of spontaneous emission power spectrum in the
vicinity of the upper synchronism frequency 100~GHz is drawn in
Fig. 2.a. The power growth along the wiggler as a function of the
wiggling periods $N_w$ is described in Fig. 2.b. In the first few
periods, the mutual interaction between the electromagnetic
radiation and the electron beam is small and the power
amplification is low. Within this stage, the spontaneous radiation
power increases proportional to $N_w^2$. An exponential growth of
SASE is inspected later after passing a sufficient number of
periods, revealing that the interaction enters to the high gain
regime, until saturation occurs when arriving to the non-linear
regime of the FEL operation. Fig. 3 describes the power evolution
in the case of grazing.

\noindent
{\bf Acknowledgments}

The research of the second author (Yu.~L.) was supported in part by
the Center of Scientific Absorption of the Ministry of Absorption,
State of Israel.


\clearpage
\def\baselinestretch{1.1}

\begin{table}
\caption{Operational parameters of millimeter wave free-electron maser.}
\label{table}
\begin{center}
\begin{tabular}{ll} 
\multicolumn{2}{l}{\underline{Accelerator}} \\
  Electron beam energy:  & $E_k$=1$\div$3~MeV \\
  Electron beam current: & $I_0$=1~A
\\[2mm]
\multicolumn{2}{l}{\underline{Wiggler}} \\
  Magnetic induction:    & $B_w$=2000~G \\
  Period:                & $\lambda_w$=4.444~cm    \\[2mm]
\multicolumn{2}{l}{\underline{Waveguide}}  \\
  Rectangular waveguide: &  1.01~cm $\times$ 0.9005~cm \\
  Mode:                  &  $TE_{01}$   \\  
\end{tabular}
\end{center}
\end{table}


\begin{figure}
\centerline{
\includegraphics[width=1.0\textwidth,angle=0]{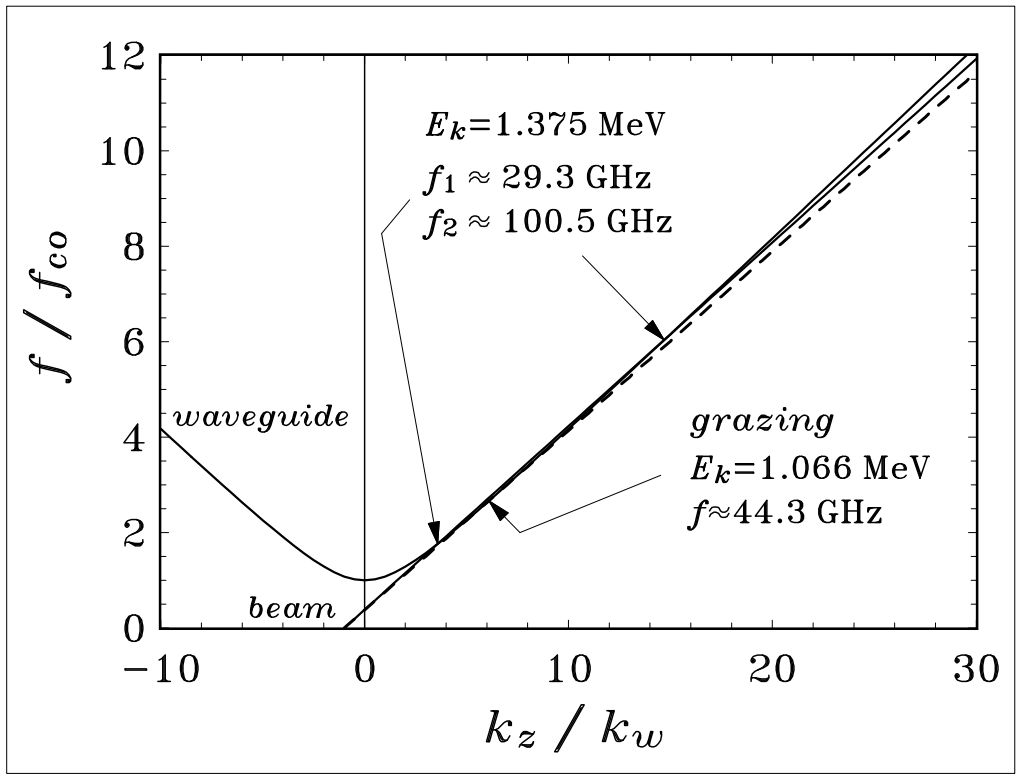}
                   }

\caption{FEM dispersion curves.}
 \label{dispersion}
\end{figure}


\begin{figure}[p]
\noindent
\centerline{
\includegraphics[height=0.47\textheight,angle=0]{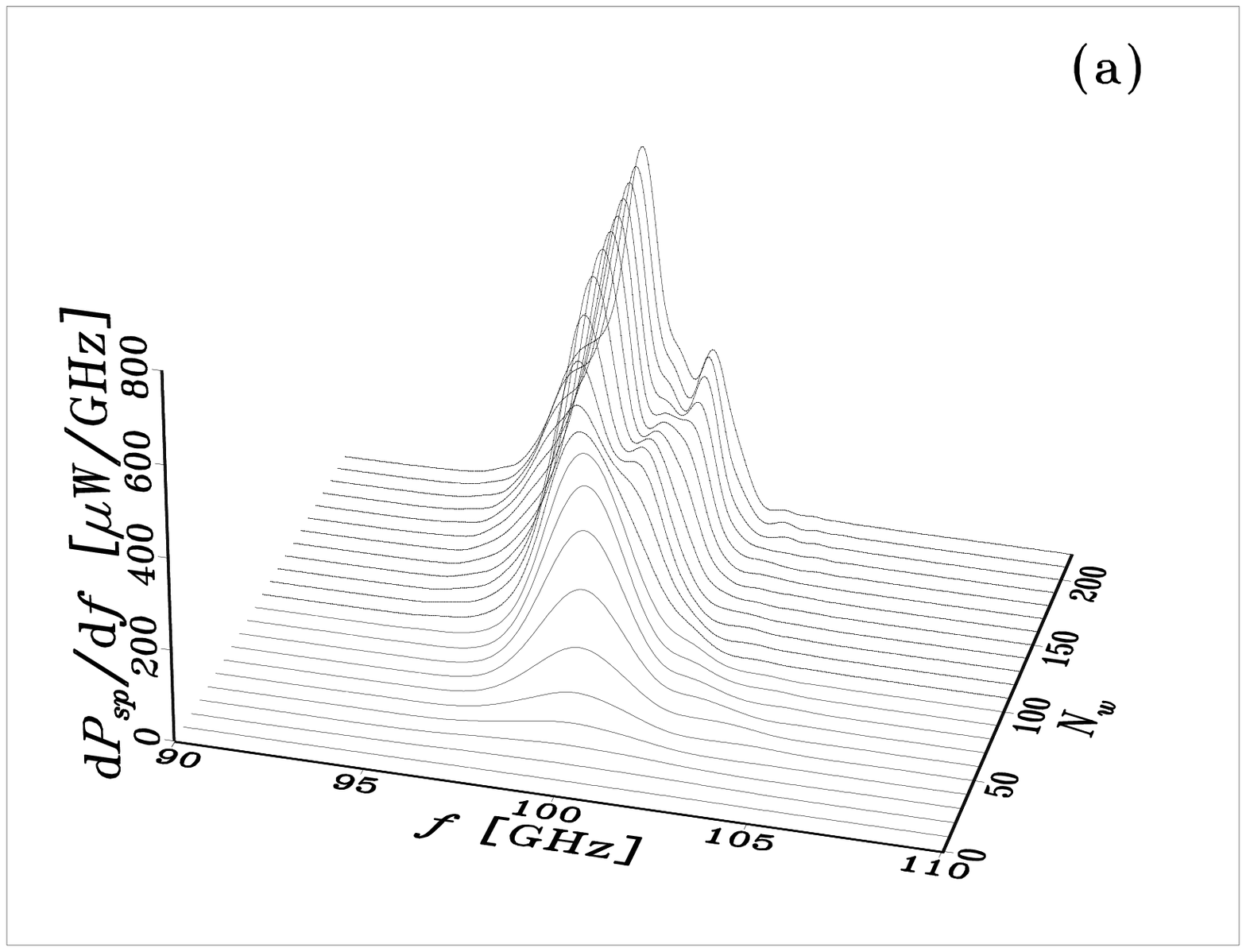}
                   }
\centerline{
\includegraphics[height=0.47\textheight,angle=0]{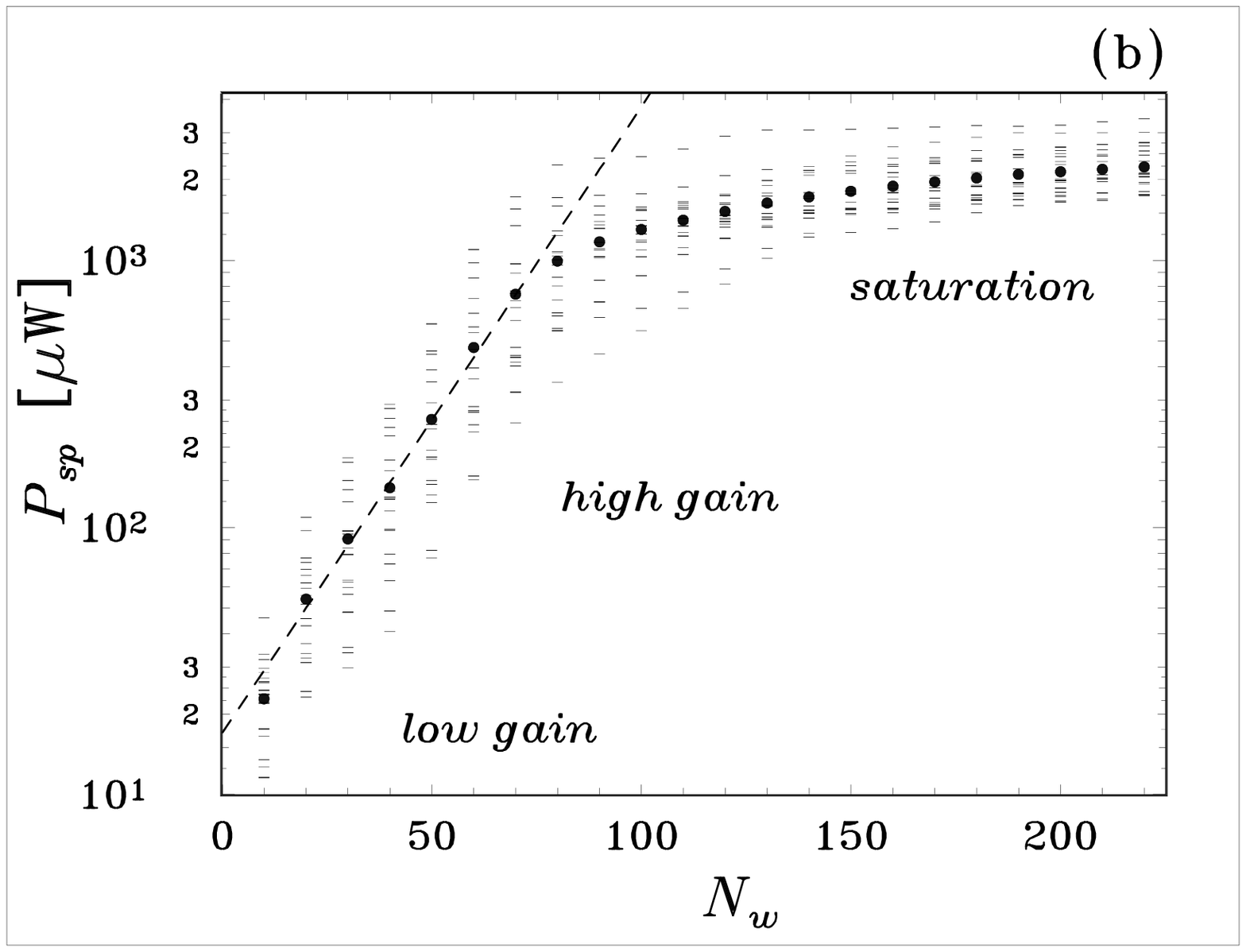}
                   }

\caption{
Evolution of spontaneous emission: (a) spectrum and (b) power
along the wiggler. The beam energy is $E_k$=1.375~MeV.
         }
\label{SASE_E1.375}
\end{figure}

%


\clearpage

\noindent
{\large\bf
\underline{Title Sheet}
}\\

\begin{itemize}
\item
\underline{Title of Paper}:
Space-frequency model of amplified spontaneous emission and super-radiance
in free electron laser operating in the linear and non-linear
regimes\\

\item
\underline{Author Name(s)}:
Yosef Pinhasi, Yuri Lurie, Asher Yahalom, and\\ Amir Abramovich
\\

\item
\underline{Author Affiliation(s)}:
The College of Judea and Samaria, Dept. of Electrical and Electronic Engineering
--- Faculty of Engineering, P.O. Box 3, Ariel 44837, Israel
\\

\item
\underline{Requested Proceedings}: Refereed

\item
\underline{Unique Session ID}: Tu-O-05

\item
\underline{Classification Codes}:
41.60.-m, 41.60.Cr, 52.75.Ms
\\

\item
\underline{Keywords}:
free electron laser, spontaneous and super-radiant emission, SASE, space-frequency 3D model
\\

\item
\underline{Corresponding Author Information}:
\begin{itemize}
\item
\underline{Full Name:} Yosef Pinhasi

\item
\underline{Postal Address}:
The College of Judea and Samaria, Dept. of Electrical and Electronic
Engineering --- Faculty of Engineering, P.O. Box 3, Ariel 44837, Israel

\item
\underline{Email Address}:
yosip@eng.tau.ac.il

\item
\underline{Telephone}:
972 - 3 - 9066 272

\item
\underline{Fax}: 972 - 3 - 9066 238
\end{itemize}
\end{itemize}

\end{document}